\documentclass[openacc]{rstransa}

\begin{document}

\title{
The fluid mechanics of poohsticks
}

\author{Julyan H. E. Cartwright$^{1,2}$, 
Oreste Piro$^{3}$
}

\address{$^{1}$Instituto Andaluz de Ciencias de la Tierra, CSIC--Universidad de Granada, E-18100 Armilla, Granada, Spain \\
$^{2}$Instituto Carlos I de F\'{\i}sica Te\'orica y Computacional, Universidad de Granada, E-18071 Granada, Spain \\
$^{3}$Departament de F\'{i}sica, Universitat de les Illes Balears, E-07071 Palma de Mallorca, Spain

}

\subject{fluid mechanics; dynamical systems}

\keywords{Maxey--Riley equation, Basset--Boussinesq--Oseen  equation, poohsticks}

\corres{Julyan Cartwright\\
\email{julyan.cartwright@csic.es} \\
Oreste Piro \\
\email{oreste.piro@uib.es}
}

\begin{abstract}
2019 is the bicentenary of George Gabriel Stokes, who in 1851 described the drag --- Stokes drag --- on a body moving immersed in a fluid,  and 2020 is the centenary of Christopher Robin Milne, for whom the game of poohsticks was invented; his father A. A. Milne's ``The House at Pooh Corner'', in which it was first described in print, appeared in 1928. So this is an apt moment to review the state of the art of the fluid mechanics of a solid body in a complex fluid flow, and one floating at the interface between two fluids in motion. Poohsticks pertains to the latter category, when the two fluids are water and air.
\end{abstract}

\begin{fmtext}

\end{fmtext}

\maketitle

\section{Poohsticks}

\begin{figure}
\centering\includegraphics[width=0.7\columnwidth]{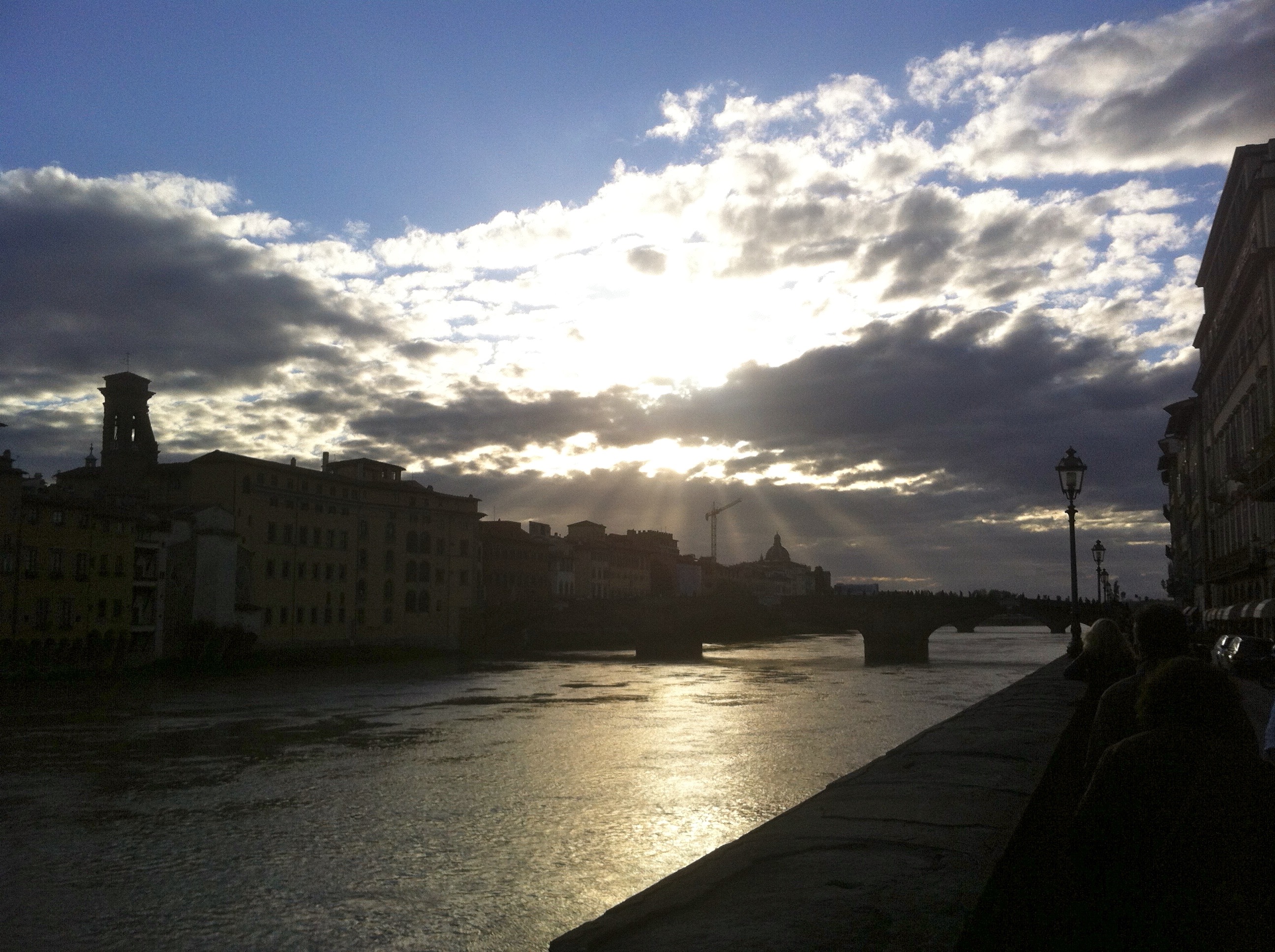}
\caption{A complex river flow made visible by the sunlight on the surface ripples emerges from under a bridge over the Arno at Florence, Italy.
}
\label{stick}
\end{figure}

The game or pastime of poohsticks is best described in the original source, A. A.  Milne's \emph{The House at Pooh Corner},
published in 1928 \cite{milne1928}. Winnie-the-Pooh
\begin{quote}
 had  just come to the bridge; and not looking where
he was going, he  tripped  over  something,  and  the  fir-cone
jerked out of his paw into the river.

       ``Bother,''  said  Pooh,  as  it floated slowly under the
bridge, and he went back to get another fir-cone  which  had  a
rhyme to it. But then he thought that he would just look at the
river instead, because it was a peaceful sort of day, so he lay
down and looked at it, and it slipped slowly away beneath him ... 
and suddenly, there was his fir-cone slipping away too.

       ``That's  funny,''  said Pooh. "I dropped it on the other
side," said Pooh, "and it came out on this side! I wonder if it
would do it again?" And he went back for some more fir-cones.

       It did. It kept on doing it. Then he dropped two in  at
once, and leant over the bridge to see which of them would come
out  first; and one of them did; but as they were both the same
size, he didn't know if it was the one which he wanted to  win,
or  the  other one. So the next time he dropped one big one and
one little one, and the big one came out first, which was  what
he  had  said  it  would  do, and the little one came out last,
which was what he had said it would do, so he had won twice ... 
and when he went home for tea, he had won thirty-six and lost
twenty-eight, which meant that he was --- that he had --- well,  you
take  twenty-eight  from  thirty-six,  and  that's what he was.
Instead of the other way round.

       And  that  was  the  beginning  of  the   game   called
Poohsticks,  which  Pooh invented, and which he and his friends
used to play on the edge of the Forest. But  they  played  with
sticks instead of fir-cones, because they were easier to mark.

       Now one day Pooh and Piglet and Rabbit and Roo were all
playing Poohsticks  together.  They had dropped their sticks in
when Rabbit said ``Go!'' and then they had hurried across to  the
other  side  of  the bridge, and now they were all leaning over
the edge, waiting to see whose stick would come out first.  But
it was a long time coming, because the river was very lazy that
day,  and  hardly seemed to mind if it didn't ever get there at
all.

       ``I can  see  mine!''  cried  Roo.  ``No,  I  can't,  it's
something  else.  Can  you see yours, Piglet? I thought I could
see mine, but I couldn't. There it is! No, it  isn't.  Can  you
see yours, Pooh?''

       ``No,'' said Pooh.
       
       ``I  expect  my  stick's  stuck,"  said Roo. "Rabbit, my
stick's stuck. Is your stick stuck, Piglet?"

       ``They always take longer than you think,'' said Rabbit.
\end{quote}
This passage not only contains a complete description of poohsticks, but also plenty of observations that speak of a scientific mind; Milne had read mathematics at Cambridge \cite{thwaite1990}. What can science tell us of a stick floating in such a fluid flow (Fig.~\ref{stick})? Some scientific questions would include, for example, ``what fluid mechanical issues determine how long it will take for a floating object to pass from one side of the bridge to another?'' and ``how and why does this time depend on the size of the floating object?''. For that understanding we need to examine equations that are now called the Maxey--Riley equations, but whose development began with George Gabriel Stokes.

\section{The motion of a solid body in a fluid}

In 1851 Stokes wrote a paper on the behaviour of a pendulum oscillating in a fluid medium, in which the famous formula for the Stokes drag of an inertial (finite-size, buoyant) body first appears \cite{stokes1851}. 
In 1885 Boussinesq \cite{boussinesq1885} added a description of the movement of a sphere settling in a quiescent fluid and in 1888 Basset \cite{basset1888} independently arrived at the same result. 
Further work was by Oseen \cite{oseen1910}, Faxen \cite{faxen1922}, and others, and at one point in the 20th century  the resulting dynamical system was known as the Basset--Boussinesq--Oseen (BBO) equation. In 1983 Maxey and Riley \cite{maxey1983} wrote down a form of equation for a small solid sphere in a moving fluid that bears their names. Gatignol \cite{gatignol1983} arrived at the same result.
Auton, Hunt and Prud'homme \cite{auton1988} pointed out in 1988 that Taylor had introduced a correction to the added mass term in a 1928 paper motivated by the motion of airships \cite{taylor1928}. The result of all this work over more than a century for the dynamics of a small rigid sphere in a moving fluid medium is today usually called the Maxey--Riley equation,
\begin{eqnarray}
&\rho_p\displaystyle\frac{d{\mathbf v}}{dt}=&
\rho_f\displaystyle\frac{D{\mathbf u}}{Dt}+(\rho_p-\rho_f){\mathbf g}
\nonumber \\
&&-\displaystyle\frac{9\nu\rho_f}{2a^2}
\left({\mathbf v}-{\mathbf u}
-\displaystyle\frac{a^2}{6}\nabla^2{\mathbf u}\right) \nonumber \\
&&-\displaystyle\frac{\rho_f}{2}\left(\displaystyle\frac{d{\mathbf v}}{dt}-
\displaystyle\frac{D}{Dt}\left[{\mathbf u}
+\displaystyle\frac{a^2}{10}\nabla^2{\mathbf u}\right]\right) 
\nonumber \\
&&-\displaystyle\frac{9\rho_f}{2a}\displaystyle\sqrt\frac{\nu}{\pi}
\displaystyle\int_{0}^{t}
\displaystyle\frac{1}{\sqrt{t-\zeta}}\displaystyle\frac{d}{d\zeta}
\left({\mathbf v}-{\mathbf u}
-\displaystyle\frac{a^2}{6}\nabla^2{\mathbf u}\right)d\zeta.
\label{eom}\end{eqnarray}
${\mathbf v}$ represents the velocity of the sphere, ${\mathbf u}$ that
of the fluid, $\rho_p$ the density of the sphere, $\rho_f$, the density of 
the fluid it displaces, $\nu$, the kinematic viscosity of the fluid, $a$, the
radius of the sphere, and $\mathbf g$, gravity. 
The derivative $D{\mathbf u}/Dt$
is along the path of a fluid element
\begin{equation}
\frac{D{\mathbf u}}{Dt}=\frac{\partial{\mathbf u}}{\partial t}
+({\mathbf u}\cdot\nabla){\mathbf u}
,\label{euler}\end{equation}
whereas the derivative $d{\mathbf u}/dt$ 
is taken along the trajectory of the sphere
\begin{equation}
\frac{d{\mathbf u}}{dt}=\frac{\partial{\mathbf u}}{\partial t}
+({\mathbf v}\cdot\nabla){\mathbf u}
.\label{lagrange}\end{equation}
The terms on the right of Eq.\ (\ref{eom})
represent respectively the force exerted by the undisturbed flow
on the particle, Archimedean buoyancy, Stokes drag, the added mass due to the boundary 
layer of fluid moving with the particle, and the Basset--Boussinesq force 
\cite{boussinesq1885,basset1888}
that depends on the history of the relative accelerations of particle and 
fluid. The terms in $a^2\nabla^2{\mathbf u}$ are the Fax\'en \cite{faxen1922} 
corrections. Equation (\ref{eom}) is derived under the assumptions that the
particle radius and its Reynolds number are small, as are the velocity gradients
around the particle. The equation is as given by Maxey and Riley \cite{maxey1983},
except for the added mass term of Taylor \cite{taylor1928} (see Auton, Hunt and Prud'homme \cite{auton1988}).

Much of the early work on a small body in a fluid, including Stokes', was undertaken to understand the forces on a pendulum moving in a fluid in order to measure as accurately as possible the acceleration due to gravity, $g$ \cite{nelson1986}.
The added mass, like the drag term, was discussed in Stokes' 1851 paper \cite{stokes1851}, where he cited Bessel's earlier work of 1828 \cite{bessel1828}. The added mass can by traced back at least to experimental work of Du Buat in 1786 \cite{dubuat1786}, and it is perhaps the earliest example of a renormalization phenomenon in physics; as has recently been pointed out by Veysey and Goldenfeld \cite{goldenfeld2007}. 

In 2000 \cite{babiano2000}, together with our colleagues Armando Babiano and Antonello Provenzale, we pointed out that from the Maxey--Riley equations one can see that even in the most favourable case of a small, neutrally buoyant particle, the particle will not always slavishly follow the fluid flow, but will on occasion bail out from following the flow and instead follow its own course for a while before rejoining the fluid flow (we subsequently studied this type of dynamical system, which is interesting in its own right, and for which we coined the term \emph{bailout embedding}, with our colleague Marcelo Magnasco \cite{cartwright2002a,cartwright2002b,cartwright2002c}). Larger particles, non-spherical particles, and particles heavier or lighter than the surrounding fluid are even less inclined to follow perfectly the fluid flow \cite{cartwright2010}. 

Let us see how things work in the most favourable case of neutral buoyancy. With this in mind, we set 
$\rho_p=\rho_f$ in Eq.\ (\ref{eom}). 
At the same time we assume that it be sufficiently
small that the Fax\'en corrections be negligible.  We also drop the
Basset--Boussinesq term to give a
minimal model \cite{russel1991,crisanti1990} with which we may perform a mathematical analysis of the
problem. If in this model there appear differences between particle
and flow trajectories, with the inclusion of further terms these discrepancies 
will remain or be enhanced.
If we rescale space, time, and velocity by scale factors $L$, $T=L/U$, and $U$, 
we arrive at the expression
\begin{equation}
\frac{d{\mathbf v}}{dt}=\frac{D{\mathbf u}}{Dt}
-{\mathrm St}^{-1}\left({\mathbf v}-{\mathbf u}\right) 
-\frac{1}{2}\left(\frac{d{\mathbf v}}{dt}-\frac{D{\mathbf u}}{Dt}\right)
,\label{neutral}\end{equation}
where ${\mathrm St}$ is the particle Stokes number
${\mathrm St}=2a^2 U/(9\nu L)=2/9\,(a/L)^2 {\mathrm Re}_f$, 
${\mathrm Re}_f$ being
the fluid Reynolds number. The assumptions involved in deriving 
Eq.\ (\ref{eom}) require that ${\mathrm St}\ll 1$. 

In the past it had been assumed that neutrally buoyant particles have trivial
dynamics \cite{crisanti1990,druzhinin1994}, 
and the mathematical argument used to back this up was that
if we make the approximation $D{\mathbf u}/Dt=d{\mathbf u}/dt$, 
which can be seen as a rescaling of the added mass, the problem becomes very 
simple
\begin{equation}
\frac{d}{dt}({\mathbf v}-{\mathbf u})=
-\frac{2}{3}\,{\mathrm St}^{-1}({\mathbf v}-{\mathbf u}) 
.\end{equation}
Thence 
\begin{equation}
{\mathbf v}-{\mathbf u}=\left({\mathbf v}_{0}
-{\mathbf u}_{0}\right)\exp(-2/3\,{\mathrm St}^{-1}\,t) 
,\end{equation}
from which we infer that even if we release the particle with a
different initial velocity ${\mathbf v}_{0}$ to that of the fluid 
${\mathbf u}_{0}$, after a transient phase the
particle velocity will match the fluid velocity, $\mathbf v=\mathbf u$, 
meaning that following this argument, a neutrally buoyant particle should be
an ideal tracer.
Although from the foregoing it would seem that neutrally buoyant particles 
represent a trivial limit to Eq.\ (\ref{eom}), this would be without taking into
account the correct approach to the problem that has $D{\mathbf u}/Dt\neq
d{\mathbf u}/dt$. If we substitute the expressions for the derivatives in
Eqs.\ (\ref{euler}) and (\ref{lagrange}) into Eq.\ (\ref{neutral}), we obtain
\begin{equation}
\frac{d}{dt}\left({\mathbf v}-{\mathbf u}\right)=
-\left(\left({\mathbf v}-{\mathbf u}\right)\cdot\nabla\right){\mathbf u}
-\frac{2}{3}\,{\mathrm St}^{-1}\left({\mathbf v}-{\mathbf u}\right) 
.\end{equation}
We may then write ${\mathbf A}={\mathbf v}-{\mathbf u}$, whence 
\begin{equation}
\frac{d{\mathbf A}}{dt}=-\left(J+\frac{2}{3}\,{\mathrm St}^{-1} I\right)
\cdot{\mathbf A}
,\label{Aeqn}\end{equation}
where $J$ is the Jacobian matrix  --- we now concentrate on
two-dimensional flows ${\mathbf u}=(u_x, u_y)$ ---
\begin{equation}
J=
\left(
\begin{array}{cc}
\partial_x u_x & \partial_y u_x \\
\partial_x u_y & \partial_y u_y
\end{array}
\right)
.\end{equation}
If we diagonalize the matrix we obtain
\begin{equation}
\frac{d{\mathbf A}_D}{dt}=
\left(
\begin{array}{cc}
\lambda-2/3\,{\mathrm St}^{-1} & 0 \\
0 & -\lambda-2/3\,{\mathrm St}^{-1}
\end{array}
\right)
\cdot{\mathbf A}_D
,\label{AD}\end{equation}
so if ${\mbox{\it Re}}(\lambda)>2/3\,{\mathrm St}^{-1}$, ${\mathbf A}_D$ may 
grow exponentially. Now $\lambda$ satisfies $\mathrm{det}(J-\lambda I)=0$, so 
$\lambda^2-{\mathrm tr}J+{\mathrm det}J=0$. Since the flow is incompressible,
$\partial_x u_x+\partial_y u_y=\mathrm{tr}J=0$,
thence $-\lambda^2=\mathrm{det}J$. Given squared vorticity  
$\omega^2=(\partial_x u_y-\partial_y u_x)^2$, 
and squared strain $s^2=s_1^2+s_2^2$, where the normal component is
$s_1=\partial_x u_x-\partial_y u_y$ and the shear component is
$s_2=\partial_y u_x+\partial_x u_y$, we may write
\begin{equation}
Q=\lambda^2=-{\mathrm det}J=(s^2-\omega^2)/4
,\end{equation}
where $Q$ is the Okubo--Weiss parameter \cite{okubo1970,weiss1991}.
If $Q>0$, $\lambda^2>0$, and $\lambda$ is real, so deformation dominates, as
around hyperbolic points, whereas if $Q<0$, $\lambda^2<0$, and $\lambda$ is
complex, so rotation dominates, as near elliptic points. Equation (\ref{Aeqn})
together with $d{\mathbf x}/dt={\mathbf A}+{\mathbf u}$ defines a dissipative
dynamical system 
\begin{equation} 
d\mbox{\boldmath$\xi$}/dt={\mathbf F}(\mbox{\boldmath$\xi$}) 
\label{disssys}
\end{equation} 
with constant divergence $\nabla\cdot{\mathbf F}=-4/3\,{\mathrm St}^{-1}$ in 
the four dimensional phase space $\mbox{\boldmath$\xi$}=(x,y,A_x,A_y)$, so that
while small values of ${\mathrm St}$ allow for large values of the divergence, 
large values of ${\mathrm St}$ force the divergence to be small. The Stokes
number is the relaxation time of the particle back onto the fluid trajectories 
compared to the time scale of the flow --- with larger ${\mathrm St}$, the 
particle has more independence from the fluid flow. From Eq.\ (\ref{AD}), about 
areas of the flow near to hyperbolic stagnation points with 
$Q>4/9\,{\mathrm St}^{-2}$, particle and flow trajectories separate 
exponentially.
The former analysis is to some extent heuristic. A rigorous approach based on an investigation of the invariant manifolds associated with the fixed points of the full 4D dynamical system was put forward by Sapsis and Haller \cite{sapsis2008}. It corroborates the analysis and even shows that our estimates are a lower bound for the actual magnitude of the separation. We have also shown that the addition of fluctuating forces such as thermal noise, for instance, enhances the magnitude of the phenomenon \cite{cartwright2002a,cartwright2002c,cartwright2002d}.

The most general description of the dynamics of these particles presents an
enormous richness of phenomena 
(see, for example, \cite{druzhinin1994,tanga1994,yannacopoulos1997,michaelides1997}). 
It is characteristic of the dynamics for small inertia that,
in the time-independent case, what were invariant surfaces in the 
model without inertia are transformed into spirals, due to centrifugal forces.
As a consequence, particles tend to accumulate at separatrices of the flow.
Except in the proximity of fixed points, the relative velocity of particle and 
fluid tends to zero at long times.
For large inertia, particles are no longer confined within vortices.
Stokes drag is the most important force acting in this case, so to a first 
approximation one can discard the other terms in Eq.\ (\ref{eom}) to obtain
\begin{eqnarray}
&&\displaystyle\frac{d^2x}{dt^2}=-\mu\left(\displaystyle\frac{dx}{dt}-u_x(x,y,t)\right), \nonumber \\ 
&&\displaystyle\frac{d^2y}{dt^2}=-\mu\left(\displaystyle\frac{dy}{dt}-u_y(x,y,t)\right),
\end{eqnarray}
where we are considering a horizontal layer with ${\mathbf v}=(dx/dt,dy/dt)$, 
and $\mu=9\nu\rho_f/(2a^2\rho_p)$.
This is a highly-dissipative and singular perturbation of a Hamiltonian 
system, with a four-dimensional phase space:
\begin{eqnarray}
&&\dot x=p_x, \nonumber \\
&&\dot p_x=-\mu(p_x-u_x(x,y,t)), \nonumber \\
&&\dot y=p_y, \nonumber \\
&&\dot p_y=-\mu(p_y-u_x(x,y,t)).
\end{eqnarray}
In the time-dependent case, particles tend to accumulate in caustics,
which correspond to the chaotic regions of the model without inertia.
The relative velocity fluctuates chaotically, due to macroscopic,
nonturbulent fluctuations, that act to give the particles deterministic but
Brownianlike motion.
In the opposite regime when the particles are lighter than the surrounding fluid --- such as air bubbles in water --- a complementary phenomenon occurs: centrifugal forces become centripetal and particles then accumulate in the core of the vortical structures of the flows.

\section{Floating bodies}

The application of the Maxey--Riley equation to floating bodies has been driven by oceanographic applications. One would like to know about the movement of drifter buoys \cite{lumpkin2017}, of flotsam and jetsam \cite{beron2016}, and of floating vegetation like sargassum \cite{gower2011}.
The desire to know where the missing airliner MH370 sank in 2014, given that bits of it were found washed up on a particular set of coastlines \cite{jansen2016,griffin2017,nesterov2018}; the dynamics of the drift of icebergs in shipping lanes \cite{mountain1980,smith1993,wagner2017}; and the finding that there are enormous `garbage patches' filled with  plastic rubbish floating at the centres of the great oceans \cite{sebille2012,ryan2014,lebreton2018} are all instances in which we should like to know how a floating body behaves in a fluid flow.

\begin{figure}
\centering\includegraphics[width=0.7\columnwidth]{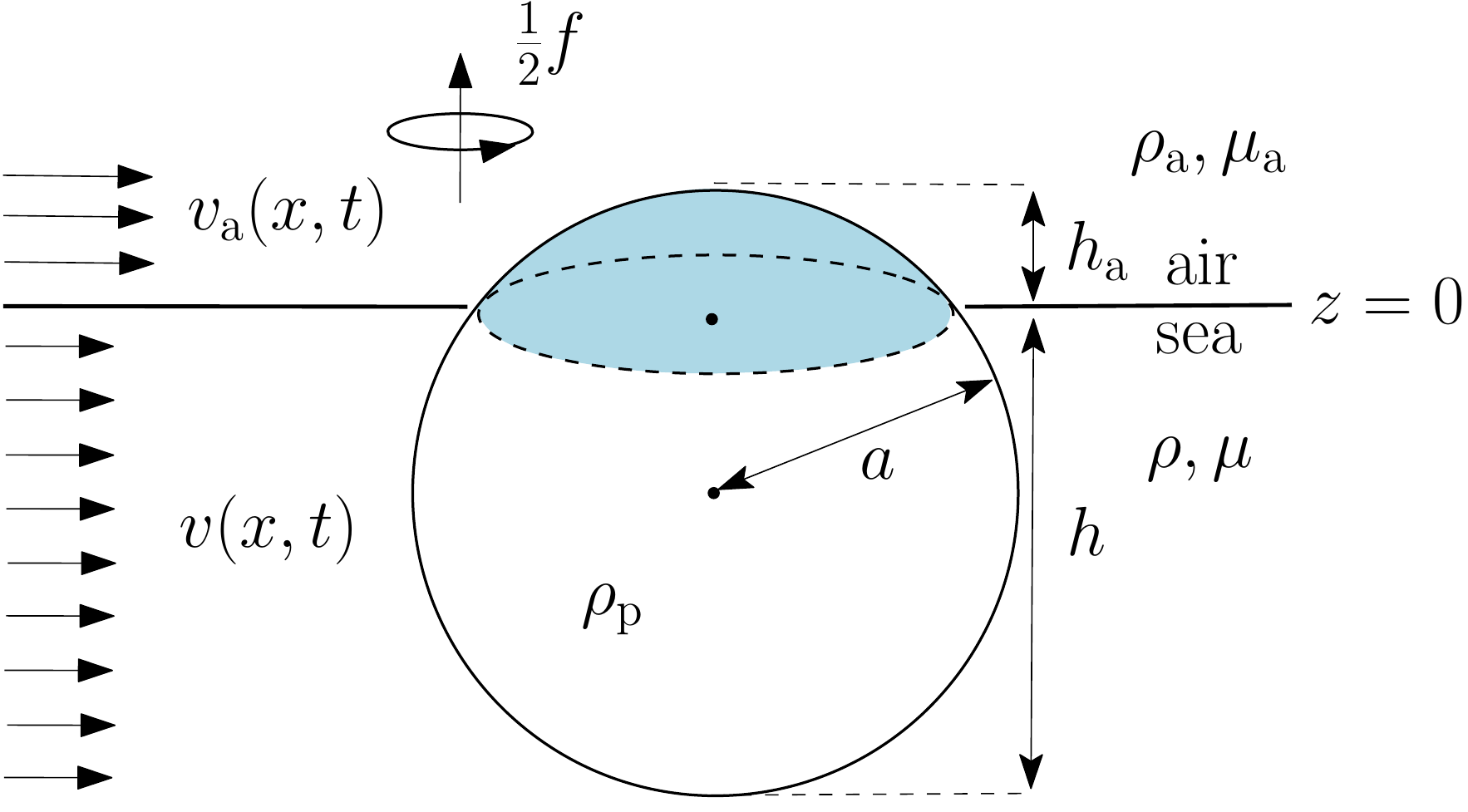}
\caption{Solid spherical particle that floats at an assumed flat interface between homogeneous water and air, and is subjected to flow, added mass, and drag forces resulting from the action of unsteady, horizontally sheared currents and winds. From Beron-Vera et al.\ \cite{beron-vera2019}.}
\label{beron}
\end{figure}

Beron-Vera and colleagues have put forward an adaptation of the Maxey--Riley equation for a sphere floating on the water surface \cite{beron-vera2019}. One can consider a floating body as one immersed in two fluid media, and each medium can have its own velocity field. This corresponds to wind over an ocean or river, for example, where the two media are  water and air; see Fig.~\ref{beron}.
For a floating sphere one obtains the expression
\begin{equation}
\dot v_p + (f+R\omega/3) v_p^\perp+\tau^{-1} v_p = R \frac{Dv}{Dt} +R(f+\omega/3)v^\perp+\tau^{-1}u
\end{equation}
where the proportional height of the spherical cap $h_a/a=\Phi$ (Fig.~\ref{beron}), and $R=(1-\Phi/2)/(1-\Phi/6)$.
Again, the case of an arbitrary shaped body is impossibly complicated, but the floating sphere can then be modified in an ad hoc fashion to try to account for shapes other than spheres using a heuristically derived shape factor.
Beron-Vera et al.\ \cite{beron-vera2019} also add Coriolis  and lift forces to the original Maxey--Riley analysis, which are present in an oceanic setting but not needed in riverine poohsticks. They note, however, that in the neutrally buoyant case, inertial particle motion is synchronized with fluid motion under the same conditions as in the original Maxey--Riley equation without Coriolis and lift effects.

\section{They always take longer than you think --- turbulent and chaotic advection}

\begin{figure}
\centering\includegraphics[width=0.7\columnwidth]{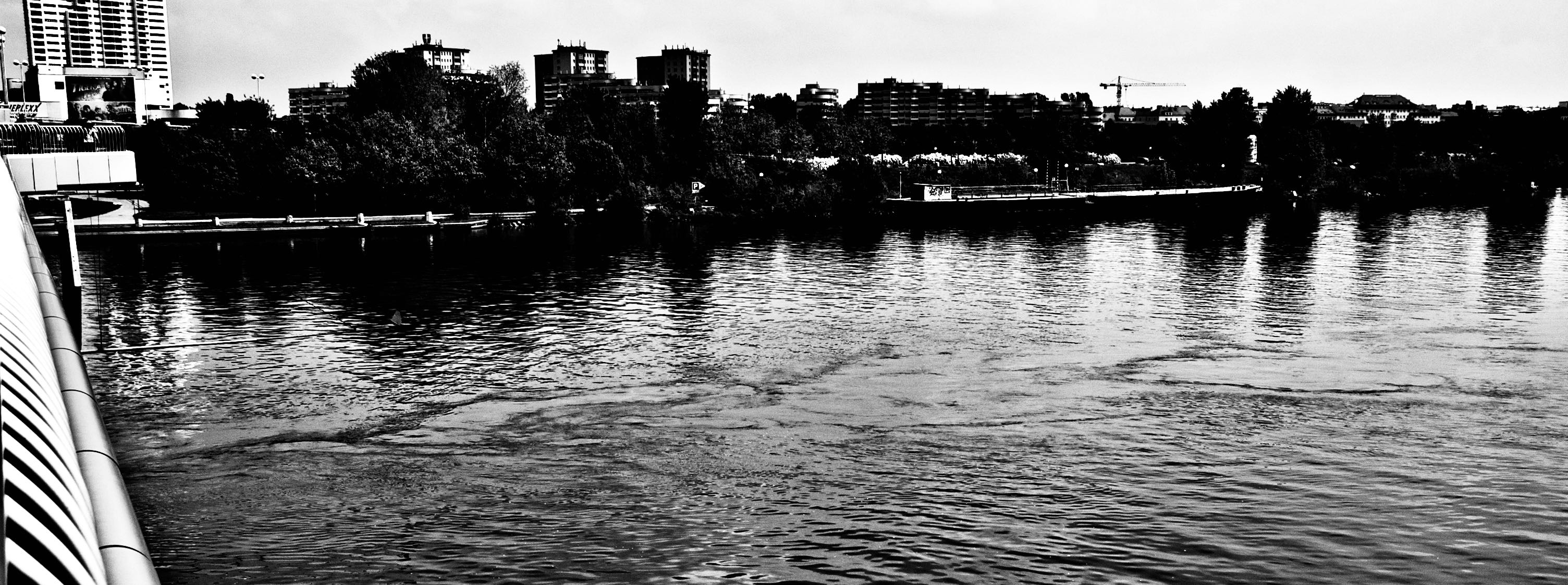}
\caption{A von K\'arm\'an vortex street stretches downstream from a bridge pillar in the Danube at Vienna, Austria.}
\label{street}
\end{figure}

\begin{figure}
\centering\includegraphics[width=0.7\columnwidth]{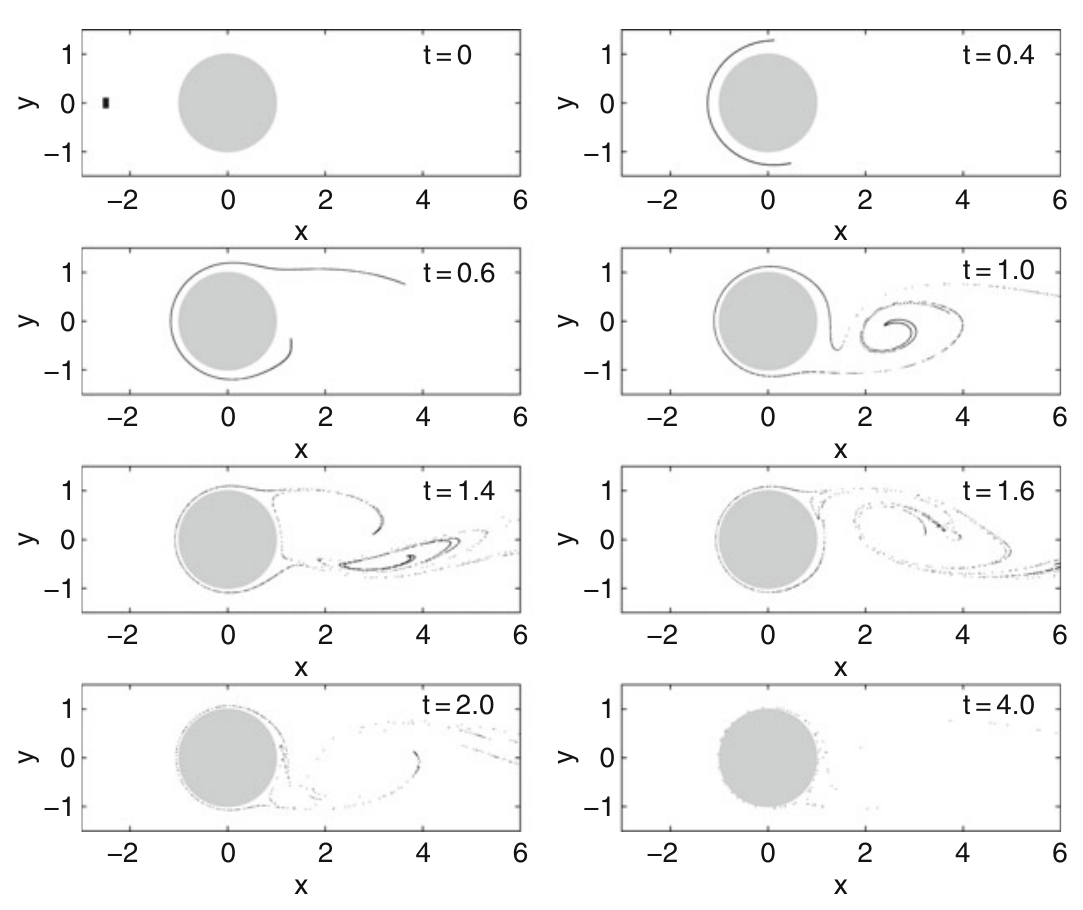}
\caption{Time evolution of a droplet of 20\,000 tracers in a von K\'arm\'an flow shown at different dimensionless times \cite{cartwright2010}.}
\label{karman}
\end{figure}

Beyond the behaviour of a solid body floating in a fluid flow that we have described up to this point,  an important  aspect of a good game of poohsticks is to use a river with an `interesting' flow field. A river in which the current simply gets stronger the closer one gets to the middle --- a Pouseuille flow profile --- would not produce a good game of poohsticks because the result would be utterly predictable: the fastest stick is the closest to the middle. It is clear from Milne's initial description that a large part of the game's interest lies in using flow fields that trap poohsticks for varying lengths of time in eddies, backflows, vortices.  When Rabbit stated ``They always take longer than you think'', he may well have been speaking of such poohstick trapping events. Bridges over streams and rivers are ideal from this point of view, since they often have pillars and pilings in the water that, as we shall describe, produce eddies.

An important influence on the complexity of the flow is the Reynolds number. Faster flows at high Reynolds number are turbulent \cite{frisch1995}, which means that they are unpredictable: the velocity field varies in both space and time. Most rivers flow fast enough that the flow is turbulent. River turbulence is of a particular kind. The surface flow in a turbulent river can to a good approximation be described as two dimensional \cite{yokosi1967,franca2015}, because the layers of fluid at different depths below the surface have very similar velocity field until one approaches the river bed. 
So one can see, for instance, vortices in the flow coming off the pillars of a bridge that are visible on the surface and  go down into the river below, because the flow field is not changing much with depth. 
This two-dimensional turbulence is seen in other instances having layers of fluid where motion perpendicular to the layers is suppressed by confinement, rotation, stratification, etc, such as in Earth's atmosphere and  oceans on a large scale. 
Two-dimensional turbulence has  special characteristics compared to its three-dimensional counterpart \cite{provenzale1999,boffetta2012}.
In 3D turbulence energy goes from large to small scales, as summarized in Richardson's famous verse \cite{richardson1922}
\begin{quote}
big whirls have little whirls
which feed on their velocity, 
and little whirls have lesser whirls 
and so on to viscosity.
\end{quote}
But in 2D turbulence it flows the other way; there is an inverse energy cascade, in the jargon.

Even slow rivers that are not turbulent, however, can give an interesting game of poohsticks. Slow flow does not have to be simple, Pouseuille-type laminar flow. Even at the low Reynolds numbers at which there is no turbulence, particles in a flow can have complicated behaviour. The appearance of complicated particle paths in low Reynolds number flow is owed to the phenomenon of chaotic advection. Chaotic advection, first described in the 1980s \cite{aref1984}, is the phenomenon of the appearance of sensitive dependence on initial conditions and complex stretching and folding processes in fluid flows. It is thus related to chaos in nonlinear dynamical systems \cite{aref2017}.

Chaotic advection is found in all sorts of slow, low Reynolds number flows. 
A river, unlike, say, a stirred cup of tea, consists of moving fluid that enters and exits a given domain; in the terminology of fluid mechanics, the water flowing under a bridge is an open, rather than a closed flow. 
There has recently been an amount of work on chaotic advection in open flows, showing how some scalars can be trapped for arbitrary lengths of time within vortices \cite{aref2017}. Milne wrote of this happening to Eeyore, who had fallen in the river, ``getting caught up by a little eddy, and turning slowly round three times'' \cite{milne1928}.

A particular instance of such eddies, important for river flows, is the von K\'arm\'an vortex street \cite{karman1963}, in which vortices are shed in an alternating fashion from either side of an obstacle in a flow. These vortex streets can often be observed downstream from bridge pilings; Fig.~\ref{street}.
Figure~\ref{karman} shows a numerical simulation of how 20\,000 points, initially close together in the black region to the left of an obstacle, are moved by the flow from left to right around the obstacle, and how they get caught within a  von K\'arm\'an vortex street to the right of the obstacle for different lengths of time, so that at the end of the simulation there are a few particles still caught within the vortices, but most have exited downstream, to the right.  Notice how the initial region becomes stretched and folded back on itself in the fluid flow, in a signature of a chaotic, nonlinear system \cite{aref2017}.

\section{Scale effects}

\begin{figure}
\centering\includegraphics[width=0.7\columnwidth]{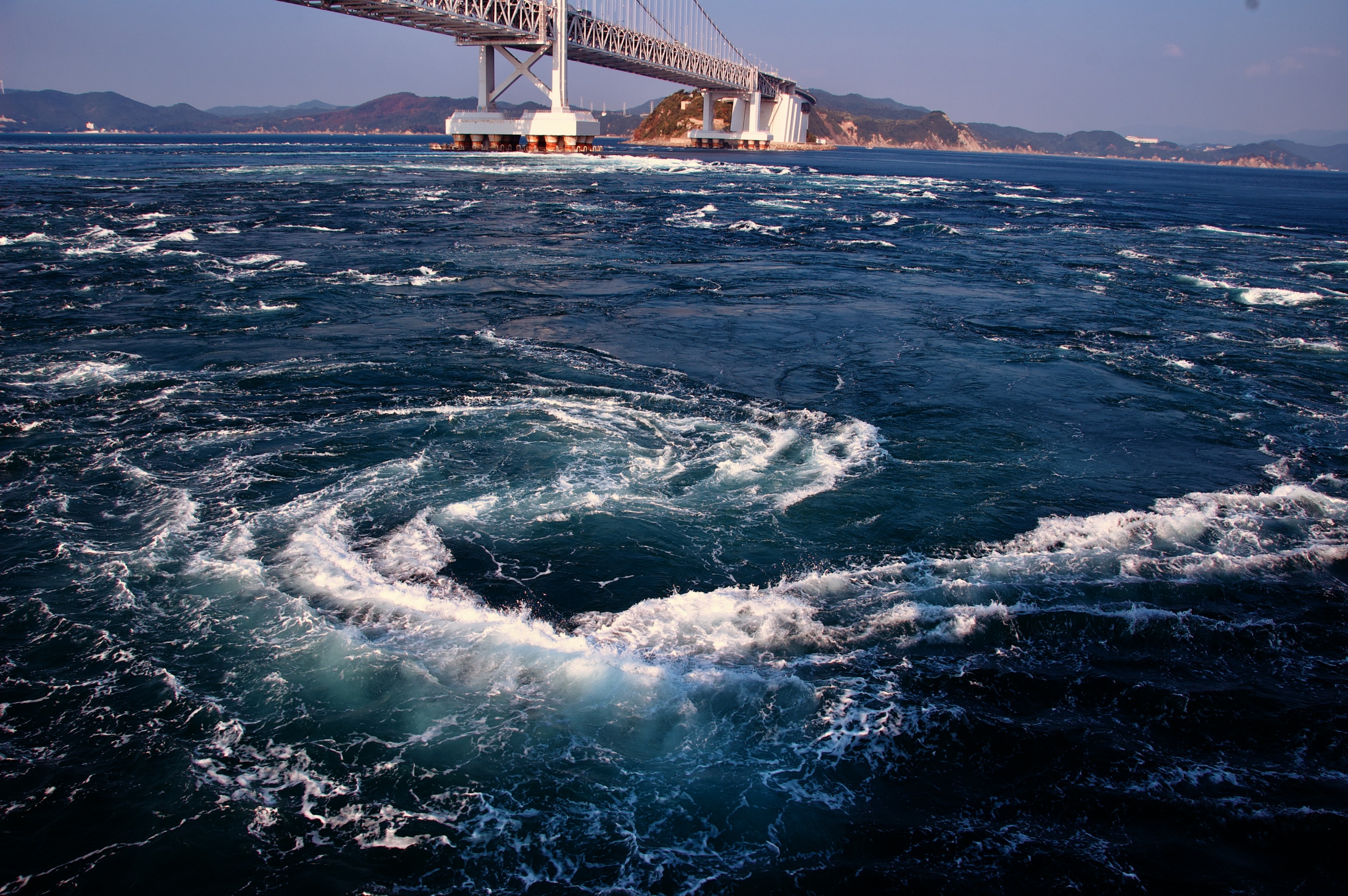}
\caption{A tidal whirlpool in the Naruto Strait between Naruto and Awaji Island, Japan.
}
\label{whirlpool}
\end{figure}

\begin{figure}
\centering\includegraphics[width=0.7\columnwidth]{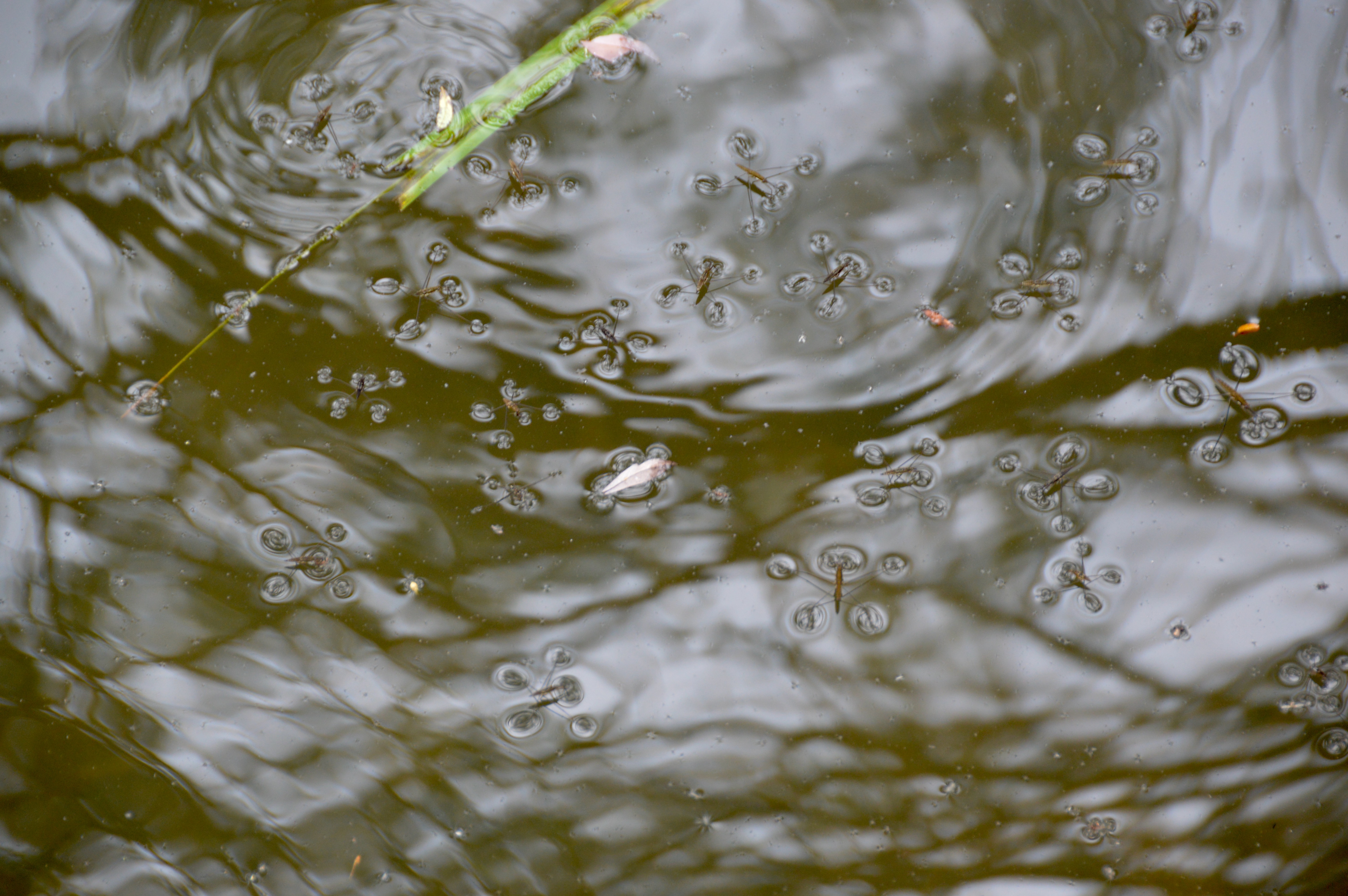}
\caption{Pond skaters on the  Duero, at Soria, Spain show surface tension effects where their legs rest on the water surface. A stick floating almost completely submerged has a different dynamics.
}
\label{skaters}
\end{figure}

Poohsticks is a game involving a centimetre to decimetre scale object in a fluid flow. This is an example of being the right size --- the poohstick is not too small or too large compared to typical length scales in the flow. Above and below this length scale, poohsticks would work out differently.
Above this length scale, one would need very large vortices in order for trapping effects to be important. That is to say, only a whirlpool like the lengendary Charybdis \cite{homer} or Maelstr\"om \cite{poe1841} can trap a ship; Fig.~\ref{whirlpool} shows a tidal whirlpool in an ocean strait. 
Below the length scale of poohsticks, on the other hand, surface tension effects start to be important \cite{keller1998}, as does flexible body dynamics \cite{burton2012}. This is the case for insects that pose on the surface of streams and rivers, like pond skaters \cite{hu2003}; Fig.~\ref{skaters}.
Milne has Eeyore explain how he escaped from being trapped in a vortex:
``You didn't think I was hooshed, did you? I dived. Pooh dropped a large stone on me, and so as not to be struck heavily on the chest, I dived and swam to the bank."
 Of course, not only living organisms that can respond actively, but also objects that alter their surface tension --- such as toy camphor boats that have been played with scientifically since the 19th century by Tomlinson, Rayleigh, and others \cite{tomlinson1862,rayleigh1889,nakata2018} --- can  move around on the water surface in ways that are quite unlike the dynamics described above.

\section{Surface waves: Stokes drift, cat turning,  and the geometric phase}

\begin{figure}
\centering\includegraphics[width=0.5\columnwidth]{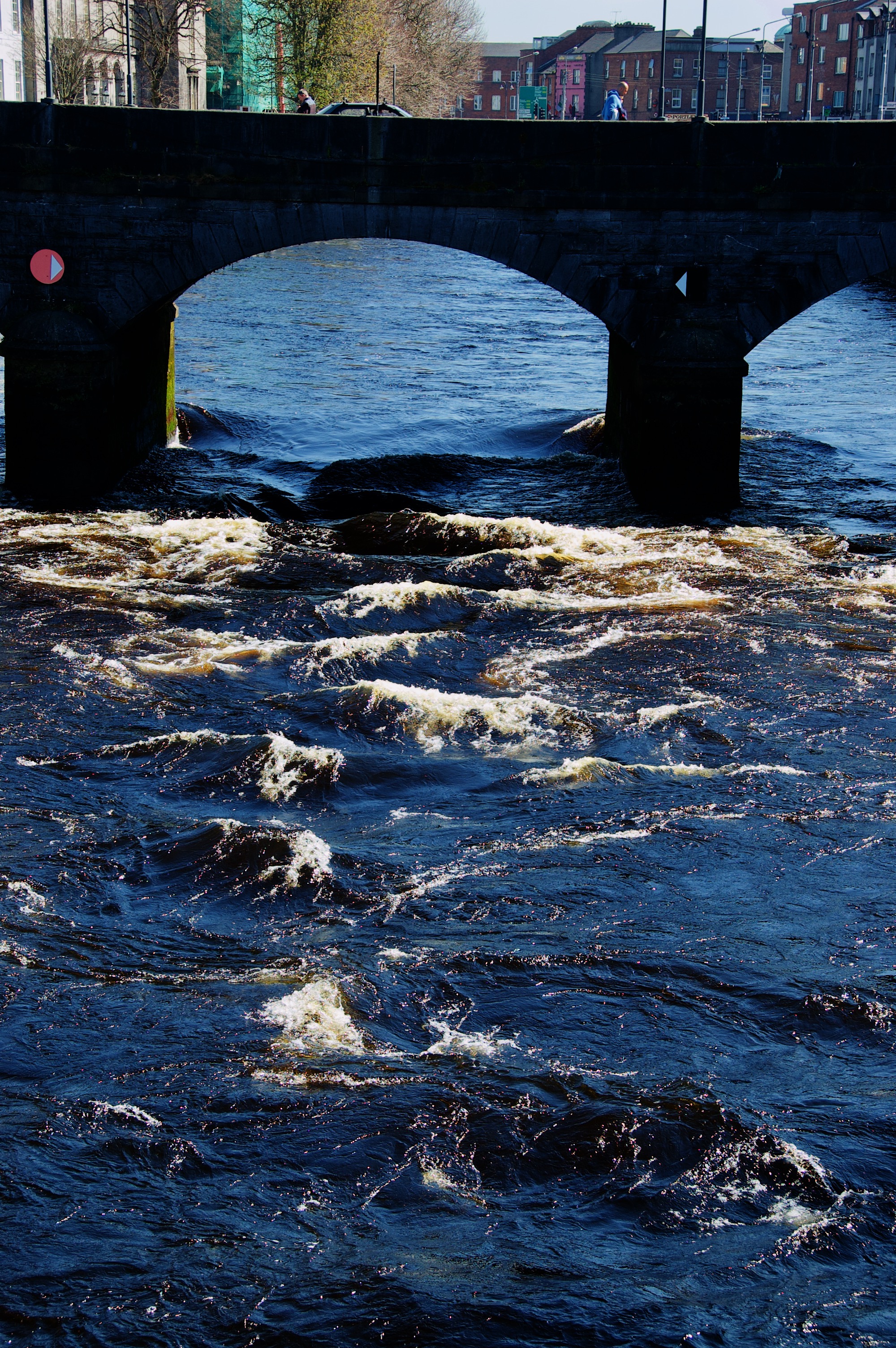}
\caption{Surface waves on the Lee at Cork, Ireland.
}
\label{waves}
\end{figure}

All of the foregoing analysis has supposed a flat interface between water and air. If there are surface waves (e.g., Fig.~\ref{waves}), there are extra effects at play. One such is Stokes drift. 
Stokes in 1847 had shown how particles in waves  may move between the arrival of one wavefront and the next in what we now call Stokes drift.
Stokes formulated the idea for deep water waves \cite{stokes1847}. Stokes drift is an example of a geometric phase or anholonomy: the failure of certain variables to return to their original values after a closed circuit in the parameters \cite{shapere1989}. 

Stokes himself, together with Maxwell, spent quite an amount of effort in the 1860s to investigate what we now know to be a further manifestation of a geometric phase in physics: cat turning. 
Maxwell himself explained the goal of the research in an 1870 letter \cite{campbell1882} to his wife
\begin{quote}
``There is a tradition in Trinity that when I was here [Trinity College, Cambridge] I discovered a method of throwing a cat so as not to light on its feet, and that I used to throw cats out of windows. I had to explain that the proper object of research was to find how quick the cat would turn round, and that the proper method was to let the cat drop on a table or bed from about two inches [we imagine that he meant to write two feet here], and that even then the cat lights on her feet.'' 
\end{quote}
Stokes' daughter Isabella recollected her father's participation in the research \cite{larmor1907}
\begin{quote}
``He was much interested, as also was Prof. Clerk Maxwell about the same time, in cat-turning, a word invented to describe the way in which a cat manages to fall upon her feet if you hold her by the four feet and drop her, back downwards, close to the floor.'' \end{quote}
In 1894 Marey published the first photographs of the phenomenon in \emph{Nature} magazine \cite{marey1894}.  But despite their cat-turning experiments and Stokes' work on particle drift in waves, Stokes and Maxwell did not discover the geometric phase. It was left to  Berry in the 1980s to point to the ubiquity of geometric phases in physics  \cite{berry3,berry2010}, and to Montgomery in the 1990s to note that cats' ability to turn as they fall is one more example of example of the failure --- in this case, desired --- of certain variables to return to their original values after a closed circuit in the parameters; a geometric phase \cite{montgomery1993}.

Although Stokes drift is a phenomenon of importance in the oceans \cite{bremer2018,beron-vera2019}, its contribution to poohsticks dynamics will generally be small in the streams and rivers in which it is typically played, unless they are like Fig.~\ref{waves}. Pooh does appear to have some understanding of the phenomenon, however, as is evidenced by his plan to rescue Eeyore from the trapping region: ``Well, if we threw stones and things into the river on one side of Eeyore, the stones would make waves, and the waves would wash him to the other side.''
Moreover, a geometric phase does come into play in a variant of the  poohsticks game in which the prediction of the orientation of the stick would play a role. How the orientation of an elongated body evolves, as the body centre of forces is advected by a vortical flow, is influenced by the presence of a geometric phase. The authors, who have previously been interested in geometric phases in discrete mappings \cite{cartwright2016} and in fluid mixing \cite{arrieta2015}, are investigating.

\section{Possibilities}

Here we have been looking into the scientific basis of a stick throwing game. It is possible that Milne might have had in mind an earlier scientific game involving both chance and throwing a stick that at one time enjoyed great popularity: Buffon's needle \cite{buffon1733}. 
If you throw a needle or stick at random onto a floor ruled with parallel lines,  such as the cracks between floorboards or tiles, from the proportion of times that the stick lands crossing a crack you can estimate $\pi$. 
We throw the 
stick --- length $l$ --- at random onto a floor --- with lines $a$ apart --- a large number of times $n$ and count the number 
of times $m$ it lands on a crack. Then $\pi$ is 
approximated by ${2ln}/(am) $; as $n$ increases, better estimates of $\pi$ are obtained.
It is probable that, as a mathematics student, Milne would have known about the Buffon's needle game and it might have been in his mind when he thought of poohsticks.
The randomness in the Buffon's needle game comes from the throwing. On the other hand, the dropping phase of a poohstick and its entry into the water do complicate matters, but probably very little, since travel times from a bridge into the water are short. Of course if a wind is blowing, that is another complication, especially if it is gusty, just like a complicated current.

We have come a long way in understanding the fluid mechanics of a solid body in a complex fluid flow in the past 170 years since Stokes' work on pendulum bobs in fluids, as we have set out to document in this review. Nonetheless, despite these advances, basic questions for which one would wish to have the answers when playing poohsticks are still open. We have a complete theory only for spherical particles, and then only for small particle Stokes number. 
Moreover, the sphere should be homogeneous in density such that the centres of mass and buoyancy coincide. 
There are some explorations of what occurs with bottom-heavy particles, like gravitactic plankton, 
 in a turbulent flow \cite{delillo2013}, but they do not integrate this with finite-size, inertial effects.  
 
 We are still at a relatively early stage in understanding the fluid mechanics of poohsticks.
Although objects larger than a trapping region will not be trapped, it is not clear how the escape of larger or longer particles smaller than the trapping region depends on the size. It is not obvious whether, all other things being equal, a larger poohstick will escape from a trapping region faster than a smaller one; whether a longer one escape faster than a shorter one, and so on. We leave the last word to Eeyore: ``Think of all the possibilities, Piglet, before you settle down to enjoy yourselves.''

\bibliographystyle{rsta}
\bibliography{pooh_sticks}

\end{document}